# Solvatochromism of (-)-Epigallocatechin 3-*O*-Gallate (EGCG) Fluorescence: Dependence on Solvent Protonicity.

**Vladislav Snitsarev [1]\*, Elena Petroff [2] and David P. Rotella [3]**

1. Montclair State University, Department of Biology, Montclair, NJ 07043, USA; SnitsarevV@montclair.edu
2. Montclair State University, Department of Biology, Montclair, NJ 07043, USA; PetroffE@montclair.edu
3. Montclair State University, Department of Chemistry and Biochemistry, Montclair, NJ 07043, USA; Margaret & Herman Sokol Institute for Pharmaceutical Life Sciences, Montclair, NJ 07043, USA; RotellaD@montclair.edu

\* Corresponding author: Tel.: (201) 250-6315; Fax: (973) 655-7047; E-mail: SnitsarevV@montclair.edu

**Abstract**

Health benefits of EGCG are well established; however, the mechanisms of EGCG action are not completely understood. In our previous study we discovered solvatochromism of EGCG fluorescence and described the dependence of fluorescence maxima on solvent polarity. We also noted that fluorescence intensity (FI) depends on solvent. The goal of this study was to gain insights into how the protonic properties of the environment affect FI. We demonstrate that 1) FI of EGCG inversely correlates with the autoprotolysis constant ($K_{ap}$) of the solvents, 2) HCl decreases, and NaOH increases FI of EGCG in water, and 3) NaOH evokes slow (~10 min time scale) transient changes in FI of EGCG. We conclude that EGCG fluorescence depends on $K_{ap}$, i.e. protonicity of the environment, that is useful for differentiating EGCG in protic aqueous environment at physiological pH~7.0÷7.4, where EGCG fluorescence is substantially quenched, from EGCG in aprotic environments, where EGCG fluorescence significantly increases. Thus, this property of EGCG fluorescence is useful for investigation of specific EGCG interactions within more aprotic protein binding sites.

## 1. Introduction

Antioxidant [1, 2], antimutagenic [3], anticancer [4-6], antiallergic [7, 8], and antiatherosclerotic [9, 10] properties of EGCG are well documented. The importance of studying EGCG-protein binding was introduced in our previous study, where we demonstrated that such a spectral property of EGCG fluorescence as the Stokes shift depended on the polarity of solvent[11]. The intensity of fluorescence, however, was not addressed. We noted, though, that the maximum FI of EGCG decreased in the following order of the solvents used: dimethyl sulfoxide (DMSO) > acetonitrile (AN) > ethanol (EtOH) >> aqueous buffer (AB). This sequence correlates with $pK_{ap}$ of the solvents, DMSO (32.7), AN (28.3), EtOH (18.9), water (14.0) (Table 1).

**Table 1.** Fluorescence intensities (FI) at $\lambda Em_{max}$ of 10 μM EGCG excited at $\lambda Ex$=275 nm in different solvents.

| Solvent | pK$_{ap}$ | FI / $\lambda Em_{max}$ (nm) |
|---|---|---|
| DMSO | 32.7 | 901.3 / 355.2<br>839.8 / 356.0<br>1047 / 356.0 |
| AN | 28.3 | 957.3 / 344.4<br>861.9 / 344.2 |
| IsoOH | 20.7 | 825.8 / 365.4<br>796.0 / 366.4 |
| EtOH | 18.9 | 624.6 / 367.0<br>605.0 / 365.8 |
| MeOH | 16.6 | 453.3 / 368.2<br>439.3 / 371.0 |
| 10% H$_2$O/90% MeOH | 14.8 | 240.7 / 372.6<br>255.5 / 370.4 |
| H$_2$O (water) | 14.0 | 15.3 / 369.8<br>16.7 / 385.8 |

pK$_{ap}$ and fluorescence intensities (FI) at $\lambda Em_{max}$ of 10 μM EGCG excited at $\lambda Ex$=275 nm are indicated for different solvents. The data are sorted in the order of decreasing pK$_{ap}$. FI/$\lambda Em_{max}$ pairs for two or three independent experiments are presented. pK$_{ap}$ are taken from [18]. See the Materials and Methods for other conditions.

FI of EGCG in AB buffered at pH 7.0 was significantly quenched compared to that in EtOH, a protic solvent that was only approximately one quarter of that in DMSO, an aprotic solvent. Two our observations, first, that FI of EGCG in AB at pH~7.0 is significantly quenched compared to EtOH, and, second, that FI of EGCG is significantly greater in aprotic solvents AN and DMSO compared to EtOH and AB at pH=7.0, point to the dependence of EGCG fluorescence on the ability of solvent to donate H$^+$. On the other hand, EGCG is ready to accept H$^+$ at the electron withdrawing carbonyl oxygen. There were two confounding parameters in our previous study[11]. First, AB was buffered with HEPES at pH=7.0, but proton transfer from nearby charged amino groups is known to quench aromatic fluorescence[12]. Second, comparing pH to K$_{ap}$ of DMSO, AN, and EtOH to pH in AB is not straightforward because K$_{ap}$ reflects the easiness of H$^+$ auto transfer or transfer to a solute if it is ready to accept H$^+$, while pH are the units of [H$^+$], or more correct, [H$_3$O$^+$] in water. This is for pure water at 25°C that K$_{ap}$=10$^{-14}$ and can be directly related to pH=7.0 (and pOH=7.0) but not for other solvents. In this study, for aqueous experiments, we used water that had been recently boiled, quickly cooled down on ice in a closed glass container, bubbled with N$_2$ and stored under N$_2$ atmosphere at room temperature to exclude the presence of CO$_2$ and O$_2$.

The effect of H$^+$ on FI depends on a particular physico-chemical system. Aniline and its derivatives are fluorescent at the neutral and alkaline pH but their fluorescence is quenched in acidic solutions when the amino group acquires H$^+$ and the anilinium ion is formed[13], whereas benzoic acid is fluorescent at acidic pH and not fluorescent at pH>5[14]. Phenol is fluorescent only in a range of pH between about 0 and 9, and quenched outside this range. In the case of EGCG, H$^+$, as an electron scavenger, can affect EGCG fluorescence through binding to the electron withdrawing carbonyl oxygen (Figure 1A) that can lead to one known and another hypothetical consequence. First, in the presence of water, a mild nucleophile, the ester bond is hydrolyzed. However, if H$^+$ is transferred to the carbonyl oxygen but there is no nucleophile strong enough to attack the carbon at the ester bond, hypothetic tautomerization of EGCG (Figure 1B) may occur. The second case may happen in organic solvents. In both cases, binding H$^+$ to the electron withdrawing carbonyl oxygen in the ester group can decrease the dipole moment of EGCG or provide a radiationless route for dissipation of the absorbed light energy [15] that leads to a decrease in FI.

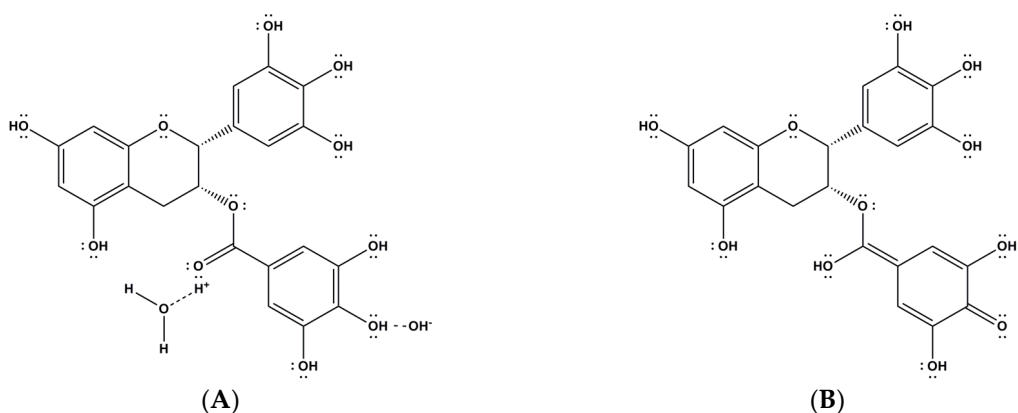

**Figure 1.** EGCG structures.

(A) EGCG is depicted with a possible coordination of $H_3O^+$ with the carbonyl oxygen in the ester group, and with a hydroxide ion to a hydroxyl in the gallate through formation of the hydrogen bonds. Though the $H_3O^+$ and $OH^-$ autoprotolytic pair is depicted, in case of other solvents it can be the corresponding autoprotolytic pair like, for instance, $CH_3OH_2^+$ and $CH_3O^-$ in case of MeOH;

(B) A possible tautomer of EGCG that can be formed by transferring $H^+$ to the carbonyl oxygen in the absence of a nucleophile able to attack the ester carbon and break the ester bond.

As noted earlier, in water, however, $H^+$-dependent withdrawal of electron density from the ester oxygen leads to hydrolysis of the ester bond, therefore the $H^+$ dependence of FI in case of EGCG is not a straightforward experiment; it is difficult to study dependence of EGCG fluorescence on $H^+$ in AB in a wide range of pH because, first, EGCG undergoes decomposition in acidic or alkaline media by hydrolysis of the ester bond and/or epimerization at the carbon atom bound to oxygen in the benzopyran ring, and, second, pH buffers are known to quench aromatic fluorescence. It is important, however, to characterize EGCG fluorescence in a less protic environment than water because binding EGCG to a binding site of an EGCG-binding protein is likely to transfer EGCG to a less protic and less hydroxylic environment. Such environment is not achievable in a standard AB, for example a HEPES-buffered salt solution (HBSS), due to the fundamental property linking $[H^+]$, $[OH^-]$, and $K_{ap}=[H^+]*[OH^-]$, because a change in $[H^+]$ always causes a reciprocal change in $[OH^-]$.

Fluorescence of EGCG can be effectively used to study binding of EGCG to proteins, and its $H^+$-dependency, if significant, may be a valuable tool because it can report transition of free EGCG from a protic aqueous environment in the cytoplasm to a non-protic environment in the binding pocket of a protein. When excited at $\lambda Ex=280$ nm, catechin, one major part of EGCG molecule, has two fluorescence emission maxima, one peak being at 314 nm and another peak ranging from 446 nm to 470 nm depending on pH of an aqueous solution [16]. Taking all together, we hypothesized that the FI of EGCG when excited at $\lambda Ex=275$ nm depends on the ability of solvent to donate $H^+$ (275 nm is $\lambda Ex_{max}$ for EGCG in EtOH).

To determine the $H^+$-dependence of EGCG fluorescence, first, we used the autoprotolytic properties of non-aqueous solvents to avoid the presence of $OH^-$. FI of EGCG in seven solvents, 1) water, 2) 10% water/90% methanol (MeOH) mixture, 3) MeOH, 4) EtOH, 5) IsoOH, 6) AN, and 7) DMSO is analyzed. Table 1 summarizes $pK_{ap}$ of the solvents used. Second, we titrated EGCG with NaOH and HCl and analyzed changes in FI in view of both the $H^+$ effects and EGCG hydrolysis happening simultaneously. Being able to distinguish FI of EGCG in these solvents would provide an important knowledge for studying EGCG binding to proteins where binding EGCG to a protein is likely to result in a transition of EGCG from an aqueous protic to a less protic environment generally resulting in changes in fluorescence intensity[15].

## 2. Results

The emission spectra of 10 μM EGCG were taken in seven solvents: DMSO, AN, IsoOH, EtOH, MeOH, 90% MeOH/10% water, and water (Figure 2A). The average FI at $\lambda Em_{max}$ from two independent runs (three for DMSO) are presented in Table 1, and plotted against $pK_{ap}$ of the solvents (Figure 2B). The regression followed monoexponential dependence.

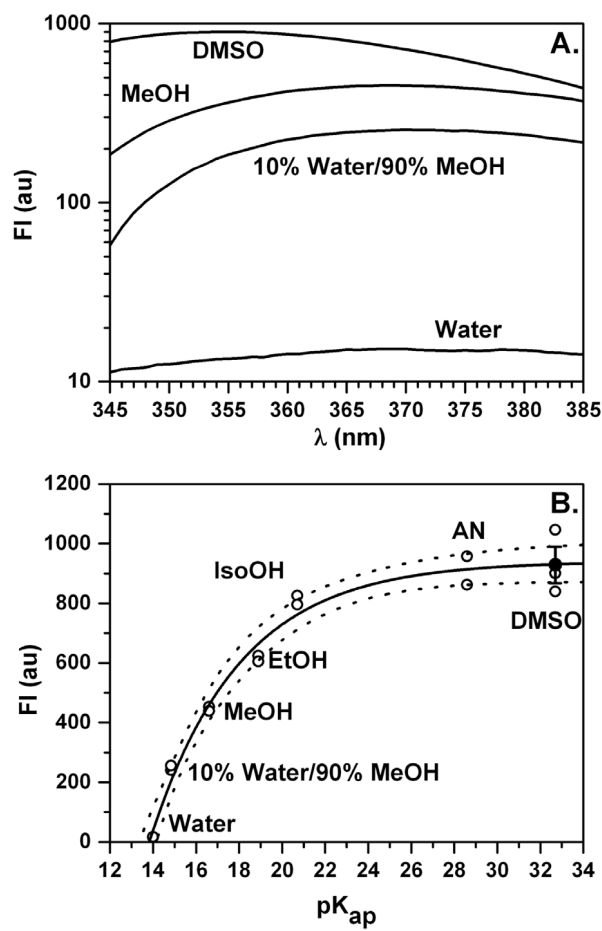

**Figure 2**. FI of EGCG in different solvents.

(A) Representative emission spectra of 10 μM EGCG excited at $\lambda Ex$=275 mm in dimethyl sulfoxide (DMSO), methanol (MeOH), 10% water/90% MeOH mixture, and water are shown. The spectra in acetonitrile (AN), isopropanol (IsoOH), and ethanol (EtOH) are not shown for clarity.

(B) FI at $\lambda Em_{max}$ is plotted against $pK_{ap}$ of the solvents. Open circles are independent runs, two for each solvent but DMSO, and three for DMSO. The closed circle is the average of three data point for DMSO±SEM. The best monoexponential fit is shown.

The emission spectra of 10 μM EGCG were compared with these immediately and 90 min after addition of 10 μM (Figure 3A), 100 μM (Figure 3B), and 1 mM (Figure 3C) NaOH. 1 μM NaOH did not cause any significant change in EGCG fluorescence for up to 90 min (Figure 3A). 100 μM NaOH decreased FI at $\lambda Em$ < ~360 nm and increased FI at $\lambda Em$ > ~360 nm immediately after NaOH addition (Figure 3B). After 90 min, the changes in FI caused by addition of 100 μM NaOH reversed, and FI increased at $\lambda Em$ < ~380 nm and decreased FI at $\lambda Em$ > ~380 nm (Figure 3B). In two independent experiments in Figure 3B, the maximum change in FI of EGCG was detected at $\lambda Em$=396 nm. 1 mM NaOH caused a decrease in FI after 90 min at all wavelengths measured (Figure 3C). At 240 nm/min scan speed, one emission scan takes 30 sec to scan from 300 nm to 420 nm, so if the changes evoked by NaOH are happening faster, the scan would have a significant time shift artifact.

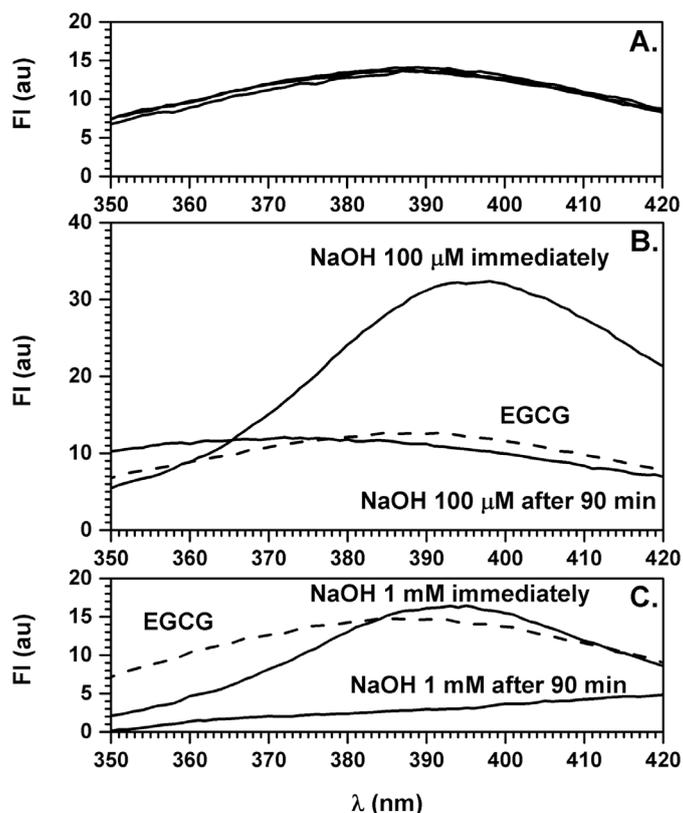

**Figure 3**. Changes in the emission spectra of 10 μM EGCG excited at λEx=275 nm in response to 10 μM, 100 μM, and 1 mM NaOH.

(A) The emission spectra of 10 μM EGCG in water, immediately after addition of 10 μM NaOH, and 90 min after were indistinguishable.

(B) EGCG spectrum in water (dashed line), immediately after addition of 100 μM NaOH, and 90 min after.

(C) EGCG spectrum in water (dashed line), immediately after addition of 1 mM NaOH, and 90 min after.

Emission scans were taken from 300 nm to 420 nm. At 240 nm/min scan speed, it takes 30 sec to complete one scan. For clarity, data from λEm=350 nm to 420 nm are shown.

To follow the behavior of FI of 10 μM EGCG in response to addition of NaOH on the second time scale, time scans at λEx=275 nm and λEm=396 nm were taken (Figure 4A). 10 μM EGCG was titrated with 1 μM, 10 μM, 100 μM, and 1 mM of NaOH (Figure 4A). The same approach was used to follow the time course of FI of 10 μM EGCG after addition of HCl. 10 μM EGCG was titrated with 1 μM, 10 μM, 100 μM, and 1 mM of HCl (Figure 4B).

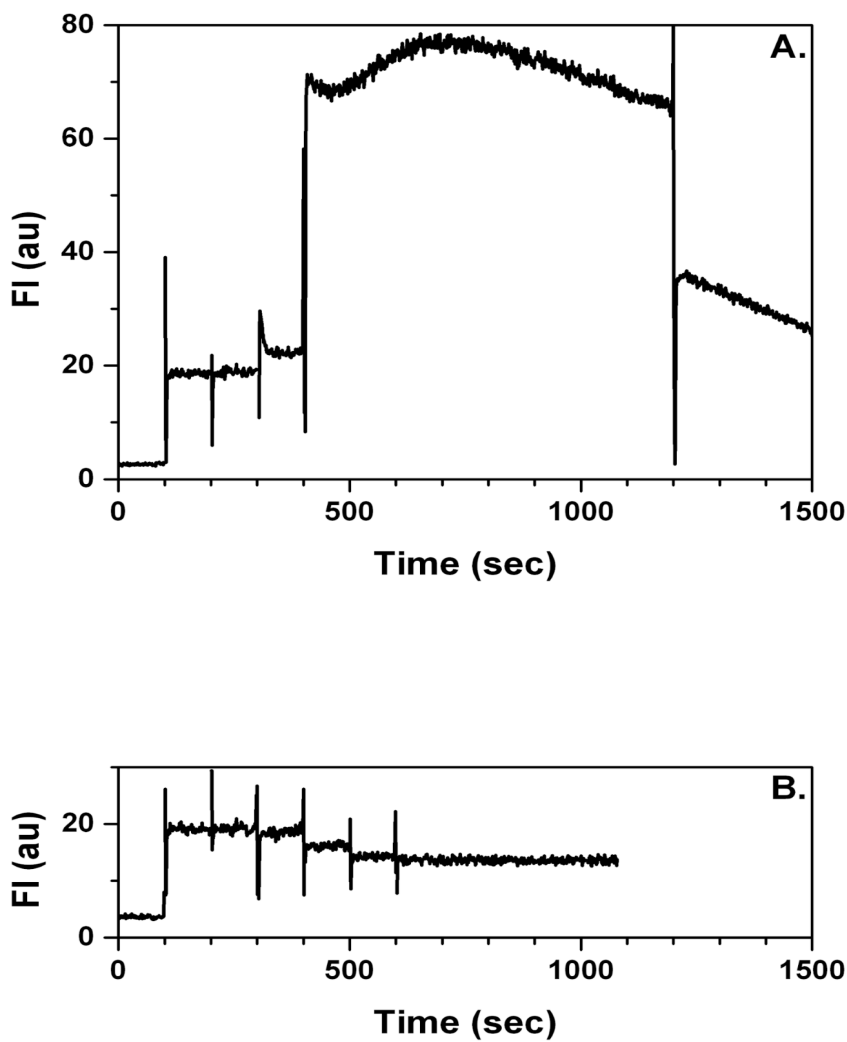

**Figure 4.** Changes in FI of 10 μM EGCG excited at λEx=275 nm and measured at λEm=396 nm in response to titration with NaOH and HCl

(A) The time course of EGCG fluorescence in response to 1 μM, 10 μM, 100 μM, and 1 mM NaOH. Between 0 sec and 100 sec, the blank measurement of the cuvette with water was taken. The additions were as following: 100" + EGCG 10 μM, 200" + NaOH 1 μM, 300" + NaOH 9 μM (10 μM total), 400" + NaOH 90 μM (100 μM total), 1200" + NaOH 900 μM (1 mM total). An average trace of two similar independent measurements is shown.

(B) The time course of EGCG fluorescence in response to 1 μM, 10 μM, 100 μM, and 1 mM HCl. Between 0 sec and 100 sec, the blank measurement of the cuvette with water was taken. The additions were as following: 100" + EGCG 10 μM, 200" + HCl 0.1 μM, 300" + HCl 0.9 μM (1 μM total), 400" + HCl 9 μM (10 μM total), 500" + HCl 90 μM (100 μM total), 600" + HCl 900 μM (1 mM total)

  2 μL additions were made directly to the cuvette via opening the light-protecting cover of the spectrofluorometer. The quartz cuvette was stirred with a Teflon-covered magnetic flea stirred at approximately 2 rotations/second.

## 3. Discussion

**FI of EGCG correlates with $K_{ap}$ of the solvents**

FI of EGCG (Figure 1) increased with an increase in $pK_{ap}$ of the solvent (Figure 2B). The data points could be fitted with a monoexponential curve. In this stud, we did not pursue to investigate the theoretical dependence of FI on $pK_{ap}$. There are a number of reasons why FI of EGCG could depend on $pK_{ap}$ of solvent. First, taking into account that EGCG is fluorescent in the UV range that points to the presence of the resonant structures with a more stable first excited state, it can be hypothesized that association of $H^+$ with the carbonyl oxygen and withdrawing of $H^+$ from gallate leads to tautomerization of EGCG (Figure 2). This hypothesis is corroborated by a fact that $pK_a$ of pyrogallol is 9.01. In solvents that are less protic than water and unable to hydrolyze EGCG such a transition seems reasonable but requires further confirmation. We'd like to make a note that in the previous paper[11], we studied EGCG in AB buffered at pH=7 with 10 mM HEPES. The low FI could be explained by the quench caused by the $H^+$ transferred from the charged amino group in HEPES, an effect known to quench aromatic fluorescence[12]. Instead of HBSS, we used here water bubbled with $N_2$ and stored under $N_2$ atmosphere at room temperature to avoid acidification with the ambient $CO_2$ and to avoid collisional quench with $O_2$ which concentration is significant in water (225 μM at 25°C) at equilibrium with atmosphere.

**Spectral response of EGCG fluorescence to NaOH**

The spectral changes of EGCG fluorescence in response to addition of NaOH demonstrated that the spectrum of 10 μM EGCG was not affected by 10 μM NaOH (Figure 3A). At 10 μM concentration, 10 μM $OH^-$ formed from dissociation of 10 μM NaOH most likely formed hydrogen bonds with a few of eight hydroxyls on each of 10 μM EGCG which is in excess (total 80 μM), so pH did not change significantly, and the ester bond remained unaffected. Hydrolysis of the ester bond would form epigallocatechin and gallic acid with different fluorescent properties. 100 μM NaOH caused first an increase in FI of EGCG, then a decrease after 90 min of exposure to NaOH (Figure 3B). Even if we assume that 10 μM EGCG with 80 hydroxyls in total can coordinate maximally 80 μM $OH^-$, $OH^-$ was in excess here and first, immediately (within 1 min) increased FI of EGCG at $\lambda Em$=396 nm. After 90 min, caused slow decrease in the FI, and, most importantly shifted the $\lambda Em_{max}$ to the shorter wavelengths that corresponds to formation of smaller size phenols, most likely epigallocatechin and gallic acid. Addition of 1 mM NaOH (Figure 3C) caused a kind of similar effect as 100 μM NaOH but of lower FI of EGCG. After 90 min, FI seen at about 300 nm in Figure 3B was significantly decreased probably to the fact that phenols do not fluoresce at pH>10.

**Time course of the EGCG FI response in response to NaOH and HCl**

Fast alkalization of 10 μM EGCG in water with 10 μM, and 100 μM but not with 1 μM NaOH resulted in a significant increase in FI in a dose-dependent manner (Figure 4A). The simplest explanation is that the carbonyl oxygen's electrons in the ester group play an active role in formation of the oscillating dipole of EGCG fluorescence, and 1 μM of $OH^-$ formed due to addition of 1 μM NaOH are coordinated through hydrogen bonding to eight hydroxyls on 10 μM EGCG that is in excess (total 80 μM). Even more, 1 μM of NaOH did not cause hydrolysis of EGCG seen at higher concentrations of NaOH making the suggestion even stronger that 1 μM of $OH^-$ ions formed from 1 μM of NaOH did not attack the ester group but were rather coordinated by the hydroxyls on EGCG.

This study opens a number of opportunities to expand further our knowledge on EGCG fluorescence such as providing theoretical explanation of the monoexponential dependence of FI on $pK_{ap}$ (Figure 2B) and elaborating on the formation of the EGCG tautomer (Figure 1B). However, we limited this report to understanding that when EGCG enters a binding pocket of a protein and undergoes transition from the aqueous cytoplasmic environment with a physiological pH around 7.0 to a less protic environment in the binding pocket of an EGCG-binding protein, FI of EGCG is expected to change significantly, namely, increase. We suggest that this significant dependence of FI of EGCG on protonicity of the environment could be useful for studying specific EGCG interactions with protein binding sites where pH and pOH are often different from aqueous environment.

## 4. Materials and Methods

### Water and HBSS

In order to avoid acidification with ambient $CO_2$ and fluorescence quench with the collisional quencher $O_2$[17], 18.2 MΩ water (Milli-Q, Millipore) was boiled, quickly cooled on ice, bubbled with $N_2$ and kept under $N_2$ atmosphere at room temperature in a closed container.

### Chemicals

All chemicals were purchased from Sigma-Aldrich Co. NMR and LC/MS confirmed the purity and structure of EGCG. EGCG was prepared as a 30 mM stock in DMSO, aliquoted with an Eppendorf Repeater® plus at 2 μL to avoid decomposition due to the freeze-thaw cycling and stored at -20°C. Before experiments, 4 μL of DMSO were added to a vial to bring EGCG concentration to 10 mM before 1/1000 dilution to the experimental concentration of 10 μM.

### Fluorescence measurements

Fluorescence emission spectra ($\lambda Ex=275$ nm) of 10 μM EGCG were measured between $\lambda Em=300$ and 420 nm with Hitachi F-7000 spectrofluorometer (Hitachi High-Technologies Co.) in a 1 cm quartz cell thermostated at 25°C. The cuvette was stirred with a Teflon-coated magnetic flea at approximately 2 turns/sec. The blank scattering spectrum of 2 mL of the solvent was taken first, then 2 μL of 10 mM EGCG was added to the cuvette with a Hamilton syringe and mixed, then the spectrum was taken again and the scattering spectrum subtracted, and all the following additions were also blank-corrected. The time course of EGCG fluorescence was followed at $\lambda Ex=275$ nm and $\lambda Em=396$ nm. The excitation and emission slit widths were 2.5 nm and 5.0 nm, respectively, PMT voltage 700 V, scan speed 240 nm/min (30 sec to complete one scan between $\lambda Em=300$ and 420 nm). The spectra and time courses were exported to and plotted with Origin 9 software (OriginLab Co).